\def\beq{\begin{equation}}
\def\eeq{\end{equation}}
\def\beqa{\begin{eqnarray}}
\def\eeqa{\end{eqnarray}}
\def\bes{\begin{split}}
\def\ees{\end{split}}

\newcommand{\bitttt}{Bi$_2$Ba$_2$Ca$_2$Cu$_3$O$_{10}$}

\newcommand{\hgot}{HgBa$_2$Ca$_{n-1}$Cu$_n$O$_{2n+2}$}

\newcommand{\tlot}{TlBa$_2$Ca$_{n-1}$Cu$_n$O$_{2n+3}$}

\newcommand{\tltt}{Tl$_2$Ba$_2$Ca$_{n-1}$Cu$_n$O$_{2n+4}$}

\newcommand{\tltttt}{Tl$_2$Ba$_2$Ca$_2$Cu$_3$O$_{10}$}

\newcommand{\cuo}{CuO$_2$}
\newcommand{\dA}{$d_{\rm A}$}
\newcommand{\ddel}{$\mathit{\Delta}\delta$}
\newcommand{\dIP}{$\delta_{\rm IP}$}
\newcommand{\dOP}{$\delta_{\rm OP}$}
\newcommand{\dP}{$\delta_{\rm p}$}

\newcommand{\nuq}{$\nu_{\rm Q}$}
\newcommand{\OA}{O$_{\rm A}$}

\newcommand{\tc} {$T_{\rm c}$}

\newcommand{\VA}{$\mathit{\Delta}\varepsilon _{\rm A}$}

\documentclass[letter,seceq]{jpsj3} 
\title{Nuclear Quadrupole Resonance Frequency in Multi-Layered Cuprates}

\author{
M. Mori$^1$, N. Afzal Shooshtary$^1$, T. Tohyama$^2$, S. Maekawa$^{1,3}$
}

\inst{
$^1$Institute for Materials Research, Tohoku University, Sendai 980-8577, Japan\\
$^2$Yukawa Institute for Theoretical Physics, Kyoto University, Kyoto 606-8502, Japan\\
$^3$JST, CREST, 5 Sanbancho, Chiyoda-ku, Tokyo, 102-0075, Japan
}

\abst{
The $^{63}$Cu nuclear quadrupole resonance (NQR) frequency, \nuq, in the multi-layered cuprates is calculated in the cluster model by the exact diagonalization method. 
The charge imbalance between the outer \cuo~plane (OP) with apical oxygen \OA~and the inner plane (IP) without \OA~in three-layered \tltttt~is estimated by comparing our results with the experimental \nuq. 
In Tl-based cuprates with more than three layers, we predict a large enhancement of the splitting of \nuq~between OP and IP by taking into account the reduction of bond length between Cu and \OA~and a resulting enhancement of the charge transfer energy.
Our results show that the NQR frequency is a useful quantity to estimate the charge imbalance in the multi-layered cuprates.
}
\kword{
nuclear quadrupole resonance, multi-layered cuprates, exact diagonalization, apical site, Madelung potential
}

\begin{document}
\maketitle
Much study has aimed the higher superconducting 
transition temperature, $T_c$. 
Regarding the high-\tc~cuprates, maximum \tc~depends on materials, e.g., \tc$\sim$40 K for La-family and \tc$\sim$90 K for Y-family, even if carrier densities are optimized. 
In particular, among hole doped cuprates, such a material dependence closely correlates with an energy difference between in-plane and apical oxygens, \VA \cite{Ohta90}. 
Actually, a spatial variation of the pairing gap due to a modulation of the apical oxygen, \OA, is observed by the scanning tunneling microscopy \cite{Slez08,Mori08}. 
In addition, it is known that disorder around the apical site is much more harmful on \tc~than other randomness \cite{Eisa06}.

On the other hand, $T_c$ of multi-layered cuprates correlates with the number of \cuo~planes in a unit cell, $n$, and shows a maximum at $n$=3\cite{Karp99,Iyo07}. 
One of the reasons why $T_c$ is suppressed for $n>$3 is the charge imbalance, i.e.,~a hole concentration in the pyramidally-coordinated-outer-planes (OP's) is different from that in the square-coordinated-inner-planes (IP's) \cite{Karp99, Trok91, Juli96, Toku00, Kote01}. 
The nuclear magnetic resonance (NMR) study has reported that the hole concentration in OP, \dOP, is larger than that in IP, \dIP\cite{Kote01}. 
The magnitude of charge imbalance, \ddel $=$ \dOP $-$ \dIP, increases with $n$, and becomes 0.10 at most for $n$=3\cite{Kote01}. 
Such a charge imbalance induces some interesting phases, e.g., two kinds of superconducting gap\cite{Toku00} and coexistence of superconducting and anti-ferromagnetic states\cite{Kote04,Muku06,Muku06JPSJ,Muku08,Mori05,Gan08}. 

To estimate the local charge density, the nuclear quadrupole resonance (NQR) frequency, \nuq, is a useful quantity \cite{Zheng95,Zheng96}, since \nuq~is determined by a local charge distribution in the ground state. 
In the NQR measurement for a three-layered \tltttt~(Tl2223)\cite{Zheng96}, the charge imbalance has been observed as two separated peaks with about 6.6 MHz difference.
Here, we note that the Knight shift, $K_c$, is also proportional to a local charge density on a measured site. 
However, one needs to examine not only the ground state but also the excited states and the neighboring spin-spin interactions, since $K_c$ is proportional to the susceptibility and the transferred hyperfine couplings\cite{Mila89}. 
In addition, one needs to measure not only $^{63}$Cu site but also $^{17}$O site to estimate the hole densities in \cuo~plane. However, the NMR measurement on $^{17}$O site is known to be rather difficult. Hence, an empirical relation between $K_c$ on $^{63}$Cu  at room temperature and the hole density estimated by the NQR measurement has been used in general\cite{Kote01}.

Previous theoretical studies on the NQR showed material dependence of \nuq~by neglecting a contribution from \OA\cite{Hanz90,Schw90,Ohta92}. 
By following Ohta {\it et al.} \cite{Ohta92}, we tried to estimate the charge imbalance observed in the NQR measurement by assuming \ddel$=$0.10 and no difference of charge transfer energy, $\Delta$, between IP and OP. 
However, our estimated value of the difference was only 2.8 MHz at most. 
The experimentally observed 6.6 MHz difference is hard to be explained without taking account of \OA\cite{note1}. 
To obtain more accurate value of charge imbalance, which correlates with \tc, one cannot ignore \OA. 

In this Letter, \nuq~on a Cu site is numerically studied in a cluster model including the 2$p_z$ orbital of \OA~by the exact diagonalization method. 
We carefully study effects of \VA~and $\Delta$ on \nuq, and show the doping dependence of \nuq~in IP and OP.
To make a link between our theoretical result and real multi-layered material, we calculate the Madelung potential of ionic model for several different series of multi-layered cuprates. 
We predict a large enhancement of the splitting of \nuq~between OP and IP in Tl-based cuprates with more than three layers.

\begin{figure}[t]
\begin{center}
\includegraphics[width=8.5cm]{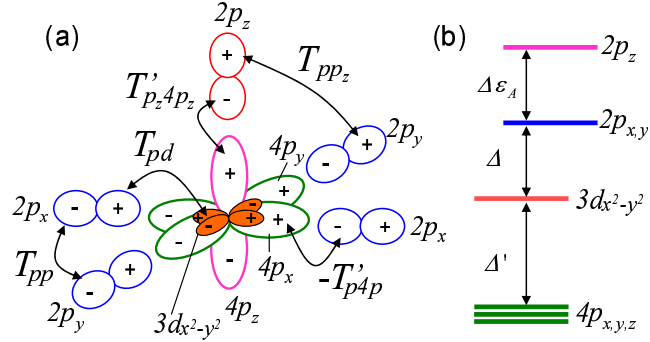}
\caption{Schematic pictures of (a) CuO$_5$ cluster with Cu 4$p_\alpha$ orbitals for OP, and (b) its energy level scheme for holes. For IP, 
CuO$_4$ cluster with Cu 4$p_{x(y)}$ orbitals is adopted, and then 2$p_z$ and 4$p_z$ orbitals are omitted\cite{Ohta92}. 
}
\label{cluster}
\end{center}
\end{figure}

The hole densities on the Cu site, which determine \nuq, are numerically calculated by the exact diagonalization method in the cluster model as shown in Fig.~\ref{cluster}(a) for OP. 
On the other hand, for IP, we adopt another cluster, in which 2$p_z$ and 4$p_z$ orbitals are omitted from the cluster in Fig.~\ref{cluster}(a)\cite{Ohta92}. 
We assume that 
the on-site Coulomb interactions  
$U_d$=8.5 eV, 
$U_{p}$=$U_{p_z}$=4.1 eV and 
$U_{4p}$=$U_{4p_z}$=2.0 eV  
are material independent\cite{Ohta91,Ohta92}, 
and 
the hopping integrals
$T_{pd}$=1.06 eV,
$T_{pp}$=0.46 eV, 
$T_{pp_z}$=0.22 eV,
$T'_{p4p}$=1.07 eV and 
$T'_{p_z4p_z}$=0.38 eV
are adjusted to the case of Tl2223 (ref.~24) by the standard bond-length dependence of 
$T_{pd}\propto d^{-4}$, $T_{pp}\propto d^{-3}$ with the parameters given in refs.~1 and 21.
The notations, $p$, $p_z$, 4$p$, 4$p_z$ and $d$, denote 2$p_{x,y}$, 2$p_z$, 4$p_{x,y}$, 4$p_z$ and $d_{x^2-y^2}$, respectively. 
Figure~\ref{cluster}(b) is the schematic picture of level separations; \VA~between $p_z$ and $p$, $\Delta$ between $p$ and $d$, and $\Delta'$ between $d$ and 4$p$. 
We assume that 4$p_{x,y}$ and 4$p_z$ are degenerate. 
For both IP and OP, $\Delta'$ is fixed to 4.8~eV\cite{Ohta92}.  
Regarding the doping rate, \dP~(p=OP, IP), we calculate two cases of hole number in the cluster: 7 and 8 holes (5 and 6 holes) in OP (IP), where 7 (5) holes are associated with the half-filling, \dOP=0 (\dIP=0). 
The case of 8 (6) holes in OP (IP) corresponds to the hole-doped case with \dOP=1 (\dIP=1).  
Since the cluster size is small, we assume that an intermediate case of \dP~could be linearly interpolated between \nuq's of \dP=0 and 1:
\beq
\nu_{\rm Q}(\delta_{\rm p})=(1-\delta_{\rm p})\nu_{\rm Q}(\delta_{\rm p}=0)+\nu_{\rm Q}(\delta_{\rm p}=1)\delta_{\rm p},  
\label{nuqdP}
\eeq
where the notations, \nuq(\dIP) and \nuq(\dOP), are used to denote the NQR frequency in IP with \dIP~and OP with \dOP, respectively. 

On a $^{63}$Cu nucleus with the quadrupole moment $Q=-0.211$ barn and the spin angular moment $I=3/2$, \nuq~in OP is calculated by 
\beqa
&&\nu _{\rm Q}=\frac{e Q}{2h}
	\Bigl\{ n_{3d_{x^2-y^2} } M_{3d}  + (4 - n_{4p_x}  - n_{4p_y} )M_{4p} \nonumber\\
&&~~~~- 2(2 - n_{4p_z} )M_{4p}\Bigr\}. 
\label{nqr}
\eeqa
The hole density on a Cu site is denoted by $n_i$ ($i$ = $3d_{x^2-y^2}, 4p_x, 4p_y, 4p_z$).  
Since the third term in eq.~(\ref{nqr}) represents the \OA~contribution to OP, 
\nuq~in IP is given by the sum of the first and the second terms. 
We use the values $M_{3d}$=$-47\times10^{21}$ V/m$^2$ and $M_{4p}$=125$\times10^{21}$ V/m$^2$\cite{Ohta92}. 
The $M_{4p}$-terms in eq.~(\ref{nqr}) arise from 4$p_\alpha$ ($\alpha$=$x$, $y$, and $z$) orbitals on a Cu site, which are partial-wave components of the 2$p_\alpha$ wave function in the neighboring O sites\cite{Schw90,Ohta92}. 
The $z$-axis is perpendicular to the plane, and the isotropy of $x$ and $y$ is assumed. 

\begin{figure}[t]
\begin{center}
\includegraphics[width=8.5cm]{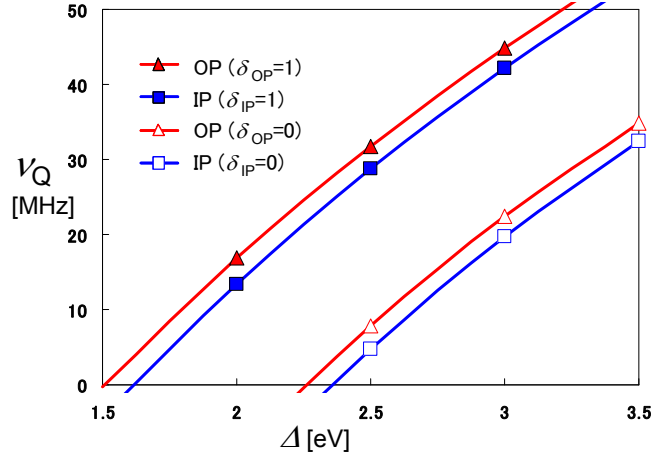}
\caption{The calculated NQR frequency \nuq~as a function of $\Delta$.
Square and triangle symbols are the results for IP and OP, and the open and closed symbols mean the undoped (\dP=0) and doped (\dP=1) cases, respectively.
These symbols are fitted by a quadratic form $\nu_{\rm Q}=A_0+A_1\Delta+A_2\Delta^2$.
}
\label{nuq}
\end{center}
\end{figure}
In Fig.~\ref{nuq}, \nuq~is plotted as a function of $\Delta$ for a fixed value of \VA=2.5~eV. We note that moderate variation of \VA~does not change the results so much. 
The symbols shown in Fig.~\ref{nuq} are fitted by a quadratic form, $\nu_{\rm Q}=A_0+A_1\Delta+A_2\Delta^2$.
We find that the values of $A_1$ and $A_2$ between OP and IP are similar for a given \dP~\cite{note2}. \nuq~increases monotonically with $\Delta$, because of the increase of $n_{3d_{x^2-y^2}}$ as reported in Ref.~21.
The large enhancement of \nuq~($\sim25$~MHz) from \dP=0 to \dP=1 is also attributed to the increase of $n_{3d_{x^2-y^2}}$.
For a given \dP, \nuq~at OP is larger by $\sim2.5$~MHz than \nuq~at IP. This is due to the contribution of the third term in eq.~(\ref{nqr}). 

It is worth emphasizing that Fig.~\ref{nuq} is available for evaluating one of the three quantities ($\nu_{\rm Q}$, $\Delta$, and \dP~(see eq.~(\ref{nuqdP})) when two of them are provided.
As an example, let us evaluate the charge imbalance $\Delta\delta$ for Tl2223 to which the 6.6~MHz difference of \nuq's between OP and IP has been reported\cite{Zheng96}.
First of all, we need to know $\Delta$ at OP and IP.
Since it is very difficult to evaluate $\Delta$ experimentally, we have to rely on theoretical results.
As will be discussed below, both $\Delta$'s at OP and IP are estimated to be $\sim$2.5~eV when $n\le 3$. Putting the estimated $\Delta$ into the fitted equation of \nuq, we find
\beq
 \nu_{\rm Q}(\delta_{\rm OP})-\nu_{\rm Q}(\delta_{\rm IP})\approx 3.0+24.0\Delta\delta.
 \label{chargeimbalance}
\eeq
This is a useful relation between the frequency difference and charge imbalance for the case of $n\le3$.
The 6.6 MHz difference of \nuq~gives $\Delta\delta$=0.15, which is slightly larger than 0.10 estimated by the empirical relation used in the NMR study\cite{Kote01}.
One possibility of this discrepancy is that the linear interpolation eq.~(\ref{nuqdP}) is too simple. 
Another one is that the empirical relation may need to be modified by considering the coordination number of oxygens around Cu site.  

\begin{figure}[t]
\begin{center}
\includegraphics[width=7.5cm]{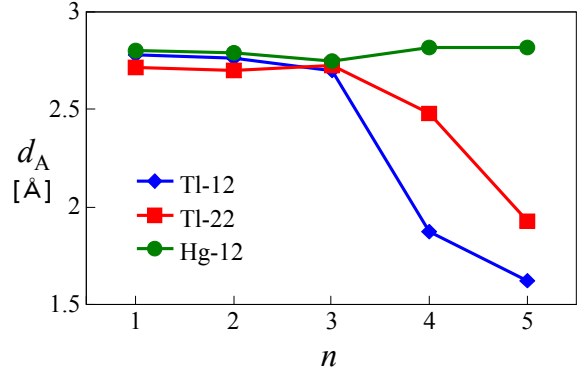}
\caption{
The bond length between Cu and \OA, \dA, versus the number of CuO$_2$ layer, $n$, for three series of multi-layered cuprates. 
The atomic coordinates are found in the following references: 
Refs. \cite{Moro88,parkin1988tcn,ihara1988nht,Naka89} for Tl-12 family,  
Refs. \cite{Izum92,Subr88,Cox88,Kiku89,Naka89} for Tl-22 family,
Refs. \cite{Haung95,hunt94,paran96,Haung95-1} for Hg-12 family. 
}
\label{dA}
\end{center}
\end{figure}

We can see in Fig.~\ref{nuq} that \nuq~sensitively changes with $\Delta$, i.e., 0.1~eV increment of $\Delta$ provides 5$\sim$6~MHz enhancement in \nuq. 
Therefore, to study the charge imbalance in the multi-layered cuprates, we need to estimate $\Delta$ in each plane. 
Below, the following three series of multi-layered cuprates are examined
(1) Tl-12 series, \tlot \cite{Moro88,parkin1988tcn,ihara1988nht,Naka89}, (\tc=48K, 110K, 123K, 112K, 107K from $n$=1 to $n$=5), 
(2) Tl-22-series, \tltt \cite{Izum92,Subr88,Cox88,Kiku89,Naka89}, 
 (\tc=95K, 118K, 125K, 112K, 105K from $n$=1 to $n$=5), 
(3) Hg-12-series, \hgot \cite{Haung95,hunt94,paran96,Haung95-1}, 
 (\tc=98K, 128K, 135K, 127K, 111K from $n$=1 to $n$=5). 
The distance between Cu and \OA, \dA, is plotted as a function of $n$ in Fig.~\ref{dA}. 
Obviously, \dA~decreases for $n>3$ except for Hg-family. 
This may be a common feature of Tl-based cuprates\cite{note3}.
The suppression of \dA~may be relevant to the suppression of \tc, since the apical oxygen position directly correlates with \VA~and the paring gap\cite{Ohta90,Mori08}. 

\begin{figure}[t]
\begin{center}
\includegraphics[width=7.5cm]{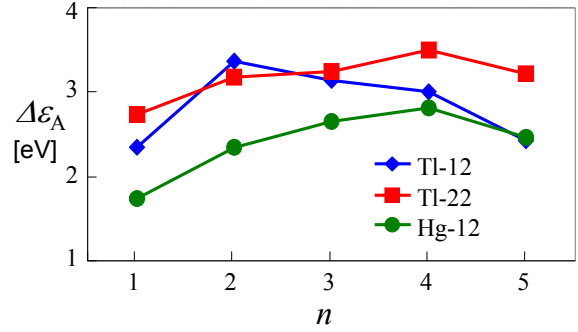}
\caption{
Potential difference between apical oxygen and in-plane oxygen, 
\VA$\equiv (V(O_{\rm A})-V({\rm O}))/\epsilon_0$, versus $n$ for three different series of multi-layered cuprates.
}
\label{va}
\end{center}
\end{figure}

To estimate \VA~and $\Delta$, we calculate the Madelung potential $V({\rm A})$ (A=Cu, \OA, and O in the plane) in the ionic model. 
In Fig.~\ref{va},  
\VA $\equiv$ ($V$(\OA)$-V$(O))/$\epsilon_0$ 
is plotted with $n$ for each family.
Note that the screening effect by the long-ranged Coulomb interaction can be effectively expressed by dividing the Madelung potentials by a constant $\epsilon_0$, whose value is assumed to be $\epsilon_0=3.5$\cite{Ohta90,Maek04}.
On the whole, \VA~does not strongly depend on $n$, while each family shows different trend. 
In Tl-12 families, \VA~slightly decreases with $n$~($>2$), while \VA~in Hg-family increases up to $n=4$. 
Since the moderate variation of \VA~shown in Fig. \ref{va} does not change \nuq~so much, we can assume \VA=2.5 eV in calculating \nuq. 

\begin{figure}[t]
\begin{center}
\includegraphics[width=8cm]{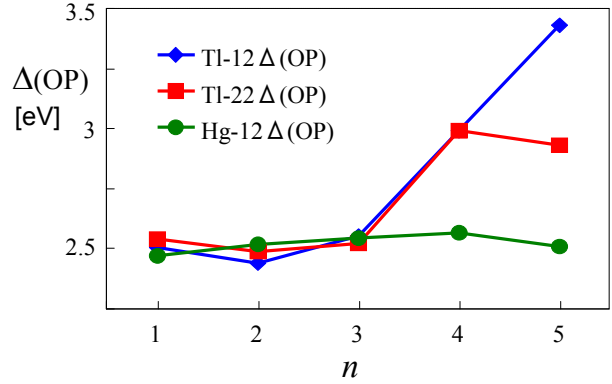}
\caption{
The charge transfer energy in the outer plane, $\Delta$(OP), versus $n$ for three series of multi-layered cuprates.  
The charge transfer energy in the inner plane, $\Delta$(IP), is almost independent of $n$ as, $\Delta$(IP)$\sim$2.5~eV.  
}
\label{D}
\end{center}
\end{figure}

On the other hand, one cannot ignore the variation of $\Delta$ shown in Fig.~\ref{D}, where $\Delta$ in OP, $\Delta$(OP), is plotted as a function of $n$. 
We used the relation, $\Delta\equiv(V({\rm O})-V({\rm Cu}))/\epsilon_0-10.9$\cite{Ohta90}. 
It is noticed that $\Delta$(OP) of Tl-families increases particularly for $n$=4 and 5. 
The deviation of Tl-family or the flatness of Hg-family must be due to their characteristic local structure as shown in Fig. \ref{dA}. 
Note that $\Delta$ in IP, $\Delta$(IP), is always about 2.5~eV independent of family. 
Let us consider Tl1245, i.e., $n$=5 in the Tl-12 family.
Taking $\Delta$(OP)=3.4~eV and $\Delta$(IP)=2.5~eV, we can predict from Fig.~\ref{nuq} a large NQR splitting, \nuq(OP)$-$\nuq(IP)$\sim$25~MHz, for the case of no carrier (\dOP=\dIP=0).
Since real Tl1245 has carriers and \dOP$>$\dIP, the NQR splitting for Tl1245 must be more enhanced than 25~MHz. 
Such a large NQR splitting has not been observed so far, but careful NQR measurements may discover a large enhancement of the NQR splitting in the Tl-families with $n$=4 and 5.
When we try to estimate the charge imbalance from observed \nuq, we should be careful about the difference of $\Delta$(OP) and $\Delta$(IP). 
For example, for Tl2234 or Tl2245 one should take $\Delta$(OP)$\approx3.0$~eV and $\Delta$(IP)$\approx2.5$~eV according to Fig.~\ref{D} in obtaining a relation between the splitting and the charge imbalance.
The relation reads $\nu_{\rm Q}(\delta_{\rm OP})-\nu_{\rm Q}(\delta_{\rm IP}) = 17.64+24.0(\delta_{\rm OP}-\delta_{\rm IP})-1.5\delta_{\rm OP}\sim 17.64+24.0\Delta\delta$. If one observed a NQR splitting, for instance, with 20~MHz, one might obtain \ddel=0.10.

To conclude, we studied the NQR frequency, \nuq, in the cluster model including the contribution from $p_z$ orbital of apical oxygen by the exact diagonalization method. 
The charge imbalance in the three-layered Tl2223 is estimated by comparing our result and the experimental \nuq. 
Our estimate on the charge imbalance \ddel$=$\dOP$-$\dIP~is 0.15, which is larger than the NMR result, \ddel$=$0.10. 
The Madelung potential in several types of multi-layered cuprates is calculated in each plane to show the $n$-dependences of energy difference between apical and in-plane oxygens, \VA, and charge transfer energy, $\Delta$.  
It is found that a large splitting of \nuq~between OP and IP can be expected in the Tl-12 and Tl-22 families.  

Recently, the angle-resolved photoemission spectroscopy (ARPES) study in \bitttt~has reported that $\delta({\rm OP})$=0.26 and $\delta({\rm IP})$=0.06, which is a quite large charge imbalance, \ddel =0.20\cite{Idet09}.  
In the ARPES study, the hole densities are estimated by the ratio of the Fermi surface volume and the first Brillouin zone. 
The ARPES result is larger than both ours and the NMR result. 
These discrepancies among theory and experiments must be solved in near future. 

We would like to thank H. Mukuda, Y. Kitaoka, A. Iyo, H. Eisaki, and A. Fujimori for useful discussions. 
This work was supported by the Grand-in-Aid for Scientific Research and Nanoscience Program of Next Generation Supercomputing Project from MEXT. 
The authors thank the Supercomputer Center, Institute for Solid State Physics, University of Tokyo for the use of facilities.


\bibliographystyle{jpsj}
\bibliography{NQR}

\begin{thebibliography}{10}

\bibitem{Ohta90}
Y.~Ohta, T.~Tohyama, and S.~Maekawa: Phys. Rev. B {\bfseries 43} (1990) 2968.

\bibitem{Slez08}
J.~A. Slezak, J.~Lee, M.~Wang, K.~McElroy, K.~Fujita, B.~M. Andersen, P.~J.
  Hirschfeld, H.~Eisaki, S.~Uchida, and J.~C. Davis: Proc. Natl. Acad. Sci. U.
  S. A. {\bfseries 105} (2008) 3203.

\bibitem{Mori08}
M.~Mori, G.~Khaliullin, T.~Tohyama, and S.~Maekawa: Phys. Rev. Lett. {\bfseries
  101} (2008) 247003.

\bibitem{Eisa06}
H.~Eisaki, N.~Kaneko, D.~Feng, A.~Damascelli, P.~Mang, K.~M. Shen, Z.-X. Shen,
  and M.~Greven: Phys. Rev. B {\bfseries 69} (2004) 064512.

\bibitem{Karp99}
M.~Karppinen and H.~Yamauchi: Mater. Sci. Eng. R. {\bfseries 26} (1999) 51.

\bibitem{Iyo07}
A.~Iyo, Y.~Tanaka, H.~Kito, Y.~Kodama, P.~Shirage, D.~Shivagan, H.~Matsuhata,
  K.~Tokiwa, and T.~Watanabe: J. Phys. Soc. Jpn. {\bfseries 76} (2007) 123702.

\bibitem{Trok91}
A.~Trokiner, L.~L. Noc, J.~Schneck, A.~M. Pougnet, R.~Mellet, J.~Primot, H.~S.
  adn Y.~M.~Gao, and S.~Aubry: Phys. Rev. B {\bfseries 44} (1991) 2426.

\bibitem{Juli96}
M.-H. Julien, P.~Carretta, M.~Horvati${\rm \acute{c}}$, C.~Berthier,
  Y.~Berthier, P.~S${\rm \acute{e}}$granan, A.~Carrington, and D.~Colson: Phys.
  Rev. Lett. {\bfseries 76} (1996) 4238.

\bibitem{Toku00}
Y.~Tokunaga, K.~Ishida, Y.~Kitaoka, K.~Asayama, K.~Tokiwa, A.~Iyo, and
  H.~Ihara: Phys. Rev. B {\bfseries 61} (2000) 9707.

\bibitem{Kote01}
H.~Kotegawa, Y.~Tokunaga, K.~Ishida, G.~q.~Zheng, Y.~Kitaoka, H.~Kito, A.~Iyo,
  K.~Tokiwa, T.~Watanabe, and H.~Ihara: Phys. Rev. B {\bfseries 64} (2001)
  064515.

\bibitem{Kote04}
H.~Kotegawa, Y.~Tokunaga, Y.~Araki, G.~q.~Zheng, Y.~Kitaoka, K.~Tokiwa, K.~Ito,
  T.~Watanabe, A.~Iyo, Y.~Tanaka, and H.~Ihara: Phys. Rev. B {\bfseries 69}
  (2004) 014501.

\bibitem{Muku06}
H.~Mukuda, M.~Abe, Y.~Araki, Y.~Kitaoka, K.~Tokiwa, T.~Watanabe, A.~Iyo,
  H.~Kito, and Y.~Tanaka: Phys. Rev. Lett. {\bfseries 96} (2006) 087001.

\bibitem{Muku06JPSJ}
H.~Mukuda, M.~Abe, S.~Shimizu, Y.~Kitaoka, A.~Iyo, Y.~Kodama, H.~Kito,
  Y.~Tanaka, K.~Tokiwa, and T.~Watanabe: J. Phys. Soc. Jpn. {\bfseries 75}
  (2006) 123702.

\bibitem{Muku08}
H.~Mukuda, Y.~Yamaguchi, S.~Shimizu, Y.~Kitaoka, P.~Shirage, and A.~Iyo: J.
  Phys. Soc. Jpn. {\bfseries 77} (2008) 124706.

\bibitem{Mori05}
M.~Mori and S.~Maekawa: Phys. Rev. Lett. {\bfseries 94} (2005) 137003.

\bibitem{Gan08}
J.~Y. Gan, M.~Mori, T.~K. Lee, and S.~Maekawa: Phys. Rev. B {\bfseries 78}
  (2008) 094504.

\bibitem{Zheng95}
G.~q.~Zheng, Y.~Kitaoka, K.~Ishida, and K.~Asayama: J. Phys. Soc. Jpn.
  {\bfseries 64} (1995) 2524.

\bibitem{Zheng96}
G.~q.~Zheng, Y.~Kitaoka, K.~Asayama, K.~Hamada, H.~Yamauchi, and S.~Tanaka:
  Physica C {\bfseries 260} (1996) 197.

\bibitem{Mila89}
F.~Mila and T.~M. Rice: Physica C {\bfseries 157} (1989) 561.

\bibitem{Hanz90}
K.~Hanzawa, F.~Komatsu, and K.~Yosida: J. Phys. Soc. Jpn. {\bfseries 59} (1990)
  3345.

\bibitem{Schw90}
K.~Schwartz, C.~Ambrosch-Draxl, and P.~Blaha: Phys. Rev. B {\bfseries 42}
  (1990) 2051.

\bibitem{Ohta92}
Y.~Ohta, W.~Koshibae, and S.~Maekawa: J. Phys. Soc. Jpn. {\bfseries 61} (1992)
  2198.

\bibitem{note1}
If we do not include a contribution from O$_\mathrm{A}$ of 3.0 MHz in eq.~(3),
  $\Delta\delta=0.275$ is necessary to reproduce 6.6 MHz difference in \nuq.

\bibitem{Ohta91}
Y.~Ohta, T.~Tohyama, and S.~Maekawa: Phys. Rev. Lett. {\bfseries 66} (1991)
  1228.

\bibitem{note2}
$A_0$=-109.527, $A_1$=58.647, $A_2$=-5.175 for IP with \dIP=0, $A_0$=-68.617,
  $A_1$=49.052, $A_2$=-4.044 for IP with \dIP=1, $A_0$=-103.463, $A_1$=57.016,
  $A_2$=-5.008 for OP with \dOP=0, and $A_0$=-62.648, $A_1$=47.482,
  $A_2$=-3.888 for OP with \dOP=1.

\bibitem{Moro88}
B.~Morosin, D.~Ginley, P.~Hlava, M.~Carr, R.~Baughman, and J.~Schirber: Physica
  C {\bfseries 152} (1988) 413.

\bibitem{parkin1988tcn}
S.~Parkin, V.~Lee, A.~Nazzal, R.~Savoy, R.~Beyers, and S.~La~Placa: Phys. Rev.
  Lett. {\bfseries 61} (1988) 750.

\bibitem{ihara1988nht}
H.~Ihara, R.~Sugise, M.~Hirabayashi, N.~Terada, M.~Jo, K.~Hayashi, A.~Negishi,
  M.~Tokumoto, Y.~Kimura, and T.~Shimomura: Nature {\bfseries 334} (1988) 510.

\bibitem{Naka89}
S.~Nakajima, M.~Kikuchi, Y.~Syono, T.~Oku, D.~Shindo, K.~Hiraga, N.~Kobayashi,
  H.~Iwasaki, and Y.~Muto: Physica C {\bfseries 158} (1989) 471.

\bibitem{Izum92}
F.~Izumi, J.~Jorgensen, Y.~Shimakawa, Y.~Kubo, T.~Manako, S.~Pei, T.~Matsumoto,
  R.~Hitterman, and Y.~Kanke: Physica. C {\bfseries 193} (1992) 426.

\bibitem{Subr88}
M.~Subramanian, J.~Calabrese, C.~Torardi, J.~Gopalakrishnan, T.~Askew,
  R.~Flippen, K.~Morrissey, U.~Chowdhry, and A.~Sleight: Nature {\bfseries 332}
  (1988) 420.

\bibitem{Cox88}
D.~Cox, C.~Torardi, M.~Subramanian, J.~Gopalakrishnan, and A.~Sleight: Phys.
  Rev. B {\bfseries 38} (1988) 6624.

\bibitem{Kiku89}
M.~Kikuchi, S.~Nakajima, Y.~Syono, K.~Hiraga, T.~Oku, D.~Shindo, K.~Hiraga,
  N.~Kobayashi, H.~Iwasaki, and Y.~Muto: Physica C {\bfseries 158} (1989) 79.

\bibitem{Haung95}
Q.~Huang, J.~W. Lynn, Q.~Xiong, and C.~W. Chu: Phys. Rev. B {\bfseries 52}
  (1995) 462.

\bibitem{hunt94}
B.~Hunter, J.~Jorgensen, J.~Wagner, P.~Radaelli, D.~Hinks, H.~Shaked,
  R.~Hitterman, and R.~Vondreele: Physica. C. Superconductivity {\bfseries 221}
  (1994) 1.

\bibitem{paran96}
M.~Paranthaman and B.~Chakoumakos: J. Solid State Chem. {\bfseries 122} (1996)
  221.

\bibitem{Haung95-1}
Q.~Huang, O.~Chmaissem, J.~Capponi, C.~Chaillout, J.~Marezio, M.and~Tholence,
  and A.~Santoro: Physica C {\bfseries 227} (1995) 1.

\bibitem{note3}
The value of \dA=2.87~\AA~for TlBa$_2$Ca$_4$Cu$_5$O$_{13}$, which is different
  from the data in ref.~29 has been reported; M.T. Weller, M.J. Pack, C.S.
  Knee, D.M. Ogborne, and A. Gormezano: Physica C {\bf 282} (1997) 849. We hope
  precise measurements of \dA~in the near future.

\bibitem{Maek04}
S.~Maekawa, T.~Tohyama, S.~Barnes, S.~Ishihara, W.~Koshibae, and G.~Khaliullin:
  {\em Physics of Transition Metal Oxides} (Springer Verlag, Berlin, 2004).

\bibitem{Idet09}
S.~Ideta, K.~Takashima, M.~Hashimoto, T.~Yoshida, A.~Fujimori, H.~Anzai,
  T.~Fujita, Y.~Nakashima, A.~Ino, M.~Arita, H.~Namatame, M.~Taniguchi, K.~Ono,
  M.~Kubota, D.~H. Lu, Z.-X. Shen, K.~M. Kojima, and S.~Uchida:
  cond-mat/0905.1223.

\end{thebibliography}

\end{document}